\documentclass[11pt]{article}
\usepackage{fullpage,epsf,amssymb,amsthm,amsfonts,amsmath,latexsym,array,graphicx,extarrows,mathtools,mathstyle,mathrsfs,slashed,subcaption,amsbsy,csquotes,accents}
\usepackage[normalem]{ulem}
\usepackage{authblk}
\usepackage{cite}
\usepackage{color}    
\usepackage{float}
\usepackage{hyperref}
\usepackage[export]{adjustbox}
\usepackage{hhline}

\numberwithin{equation}{section}

\title{\bf{On the (in)consistency of perturbation theory at finite temperature}} 

\author[1]{Peter Lowdon}
\author[1,2]{Owe Philipsen}

\affil[1]{{\scriptsize Institut f\"{u}r Theoretische Physik, Goethe-Universit\"{a}t, Max-von-Laue-Str. 1,  60438 Frankfurt am Main, Germany}}
\affil[2]{{\scriptsize John von Neumann Institute for Computing (NIC) at GSI, Planckstr.\ 1, 64291 Darmstadt, Germany}}

\date{}
\begin{document}
\maketitle

\begin{abstract}
\noindent
A well-known difficulty of perturbative approaches to quantum field theory at finite temperature is the necessity to address theoretical constraints that are not present in the vacuum theory. In this work, we use lattice simulations of scalar correlation functions in massive $\phi^{4}$ theory to analyse the extent to which these constraints affect the perturbative predictions. We find that the standard perturbative predictions deteriorate even in the absence of infrared divergences at relatively low temperatures, and that this is directly connected to the analytic structure of the propagators used in the expansion. This suggests that the incorporation of non-perturbative thermal effects in the propagators is essential for a consistent perturbative formulation of scalar quantum field theories at finite temperature. By utilising the spectral constraints imposed on finite-temperature correlation functions, we explore how these effects manifest themselves in the lattice data, and discuss why the presence of distinct thermoparticle excitations provides a potential resolution to these issues.
\end{abstract}

\newpage

\section{Introduction}
\label{intro}

The inclusion of a finite temperature $T$ in 3+1-dimensional quantum field theories (QFTs) is known to introduce complications that have a significant effect on their perturbative treatment. In loop corrections to the self-energies or the pressure, powers of the coupling generically appear together with powers of $T/m$. In theories with massless bosons, infrared divergences occur for $T>0$ which cause the perturbative expansion to break down at a fixed loop order. These apparent divergences are expected to be cured by the infrared dynamics of the theory, and in practice are treated by resumming classes of diagrams to all loop orders~\cite{Kapusta:2006pm,Bellac:2011kqa,Laine:2016hma}. For non-abelian theories, such as in the symmetric electroweak phase or in Quantum Chromodynamics (QCD), dynamically generated masses $m\sim gT,g^2T$ cancel the gauge coupling in the effective dimensionless expansion parameter $\sim g^2T/m$ from an observable-specific loop order upwards, rendering these thermal theories completely non-perturbative beyond some low order in the gauge coupling~\cite{Linde:1980ts,Gross:1980br}. Even in QFTs for which this does not occur, such as scalar theories, the resummed perturbative series suffers poor convergence properties~\cite{Kapusta:2006pm,Bellac:2011kqa,Laine:2016hma,Blaizot:2003tw,Andersen:2004fp}. Attempts to resolve these infrared problems include the use of effective field theory~\cite{Braaten:1994na}, systematic computations of higher-loop diagrams~\cite{Kajantie:2002wa,Gynther:2007bw}, and various reorganisations of the perturbation expansion such as optimised infinite resummations~\cite{Karsch:1997gj,Andersen:2000yj,Andersen:2008bz,Chiku:1998kd,Braaten:1989mz}, and variational techniques like the two-particle irreducible (2PI) formalism~\cite{Blaizot:2000fc,vanHees:2001ik,VanHees:2001pf,vanHees:2002bv}. \\

\noindent
These infrared characteristics are generally viewed as a purely technical feature of perturbation theory. However, they may in fact be a symptom of a more fundamental non-perturbative constraint that arises from a highly consequential result of finite-temperature QFT: the \textit{Narnhofer-Requardt-Thirring} (NRT) theorem~\cite{Narnhofer:1983hp}. The theorem implies that if a scattering matrix $S$ is constructed from $T>0$ states with a dispersion relation $\omega=E(\vec{p})$, where $E(\vec{p})$ is some real function, then $S$ is necessarily trivial, i.e. $S=1$. This therefore rules out the possibility of having interacting states at finite temperature with purely real dispersion relations. For scalar vacuum QFTs it is rigorously known that the large-time scattering states have an on-shell Fock structure, which explains why free-field propagators appear in the perturbative series~\cite{Haag:1992hx}. But because of the NRT theorem this cannot be the case when $T>0$. In particular, this means that neither free fields, nor quasi-particle-like propagators with real poles, can form the basis of a consistent finite-temperature perturbative expansion~\cite{Landsman:1988ta}. \\

\noindent
The stark consequences of the NRT theorem arise because thermal states satisfy the Kubo-Martin-Schwinger (KMS) condition, which implies vastly different spectral properties than those in the vacuum case~\cite{Landsman:1988ta,Haag:1967sg}. These differences are a reflection of the fact that the dissipative effects of a thermal medium in equilibrium are present everywhere and at all times. These effects must therefore be taken into account in the definition of \textit{every} state, including those at large times, which ultimately form the basis of any perturbative expansion. In this sense, the NRT theorem reflects a primitive constraint which is not simply restricted to infrared regimes, but should quantitatively affect systems also at low temperatures, or with non-vanishing mass scales. At first sight the implications of the NRT theorem appear to be of a mainly formal nature. However, it has been demonstrated for scalar QFTs that perturbation theory does indeed break down if the propagators used in the expansion have a real dispersion relation~\cite{Weldon:1998bj,Weldon:1998xr,Weldon:2001vt}. This includes the standard approach of using free propagators, as well as any resummed expansion in which the (retarded) propagator has the general form
\begin{align}
G_{R}(p) = -\frac{1}{(\omega + i \epsilon)^{2} - E(\vec{p})^{2}},
\end{align}
where $E(\vec{p})$ is a real-valued function. In these specific cases the breakdown arises because the self-energy $\Pi(k)$ develops a branch point singularity at $k_{0}=E(\vec{p})$, which prohibits the computation of perturbative corrections to the propagator pole. In Ref.~\cite{Weldon:2001vt}, resummed perturbative results for massless scalar $\phi^{4}$ theory~\cite{Wang:1995qf} were used to show that this breakdown occurs explicitly at two-loop order in this case. There have been some suggestions~\cite{Landsman:1988ta,Weldon:1998xr,Bros:1992ey,Buchholz:1993kp,Bros:1995he,Bros:1996mw,Bros:2001zs,Bros:2003zs} as to how one might circumvent these finite-temperature constraints, but ultimately this remains an open question. \\

\noindent
In this study, we show by explicit calculation that the issues raised above quantitatively affect the perturbative predictions for massive real $\phi^{4}$ theory in the symmetric phase, even in the regime $0<T/m \lesssim \mathcal{O}(1)$ where infrared divergences are absent. For this purpose we compare the perturbative two-loop correlation functions with the results from lattice simulations. Since the latter contain the full dynamics of the theory, this allows one to assess the relative importance of non-perturbative effects as a function of temperature and the coupling strength. In order to avoid triviality of the fully non-perturbative theory in the continuum limit~\cite{Wilson:1973jj,Frohlich:1982tw,Luscher:1987ay} we keep a finite lattice spacing in all cases and work with lattice perturbation theory on finite volumes, which is known to be perturbatively renormalisable to all orders~\cite{Reisz:1987da,Reisz:1987hx,Reisz:1987pw,Reisz:1987px}. In this way we retain a non-trivial interacting theory which is accessible at high precision, and in addition eliminate any mixing of cutoff and finite-size effects with the systematics of perturbation theory. In the final part of this work we focus on the ideas put forward in Refs.~\cite{Bros:1992ey,Buchholz:1993kp,Bros:1995he,Bros:1996mw,Bros:2001zs,Bros:2003zs}, and demonstrate that they provide a promising direction in which to resolve the perturbative issues encountered at finite temperature. \\

\noindent
The remainder of the paper is structured as follows: in Sec.~\ref{latticePT} we provide an overview of lattice perturbation theory and its application to $\phi^{4}$ theory, in Sec.~\ref{lattice_analysis} we discuss the lattice setup, data analysis, and theoretical implications, in Sec.~\ref{PT_issues} we analyse the ideas of Refs.~\cite{Bros:1992ey,Buchholz:1993kp,Bros:1995he,Bros:1996mw,Bros:2001zs,Bros:2003zs} in the context of our results, and in Sec.~\ref{concl} we summarise our findings.

\section{Lattice $\phi^{4}$ theory for a real scalar field}
\label{latticePT}

\subsection{Lattice formulation}

As outlined in Sec.~\ref{intro}, there are strong suggestions that the conventional approach to perturbation theory at finite temperature has systematic inconsistencies which go beyond the well-known infrared issues one encounters in massless theories, or when $T/m \gg 1$. If these inconsistencies are indeed present, then the perturbative prediction should worsen as the temperature increases. In order to test this hypothesis we perform lattice simulations of two-point correlation functions in $3+1$-dimensional $\phi^{4}$ theory, and compare these data with the corresponding predictions from finite-temperature lattice perturbation theory. Before outlining the overall analysis strategy we first provide a brief overview of lattice $\phi^{4}$ theory, as detailed in Ref.~\cite{Montvay:1994cy}. We consider an isotropic, four-dimensional Euclidean spacetime lattice with spacing $a$, consisting of the set of points $\Lambda_{a} = \{ x \in a \mathbb{Z}^{4}  : 0 \leq x_{\mu} \leq a(N_{\mu}-1)\}$, with the same extent in all spatial directions, $N_{x}=N_{y}=N_{z}=N_{s}$. The real scalar field $\phi(x)$ defined on those points has periodic boundary conditions in all directions: $\phi(x_{\mu}+a\hat{n}_{\mu}N_{s})=\phi(x_{\mu})$. Since thermal states at a temperature $T=1/\beta$ are defined to have $\beta$-periodic Euclidean correlation functions in the temporal direction, by virtue of the KMS condition, the lattice correlators can be identified with those of a thermal QFT with $T=1/(aN_{\tau})$~\cite{Montvay:1994cy}. The vacuum QFT at fixed lattice spacing and spatial volume $V=L^{3}=(aN_{s})^{3}$ is then recovered in the limit $N_{\tau}\rightarrow\infty$.\\

The lattice action in $\phi^{4}$ theory is defined by
\begin{align}
S = a^{4}\sum_{x \in \Lambda_{a}} \left[ \frac{1}{2}\sum_{\mu} \Delta_{\mu}^{f}\phi(x)\Delta_{\mu}^{f}\phi(x)+ \frac{m_{0}^{2}}{2}\phi(x)^{2} +\frac{g_{0}}{4!}\phi(x)^{4} \right], 
\label{eq:latact1}
\end{align}
where $\Delta_{\mu}^{f}$ is the lattice forward derivative: $\Delta_{\mu}^{f}\phi(x)= \left[\phi(x+a\hat{n}_{\mu})-\phi(x)\right]/a$. The model has a $\mathbb{Z}_{2}$ reflection symmetry, which gets spontaneously broken for $m_{0}^{2}<0$. In this study we focus on the symmetric phase and stay clear of infrared divergences by working with bare masses $m_{0}^{2}>0$. On the lattice the Fourier transform and its inverse are given by
\begin{align}
\widetilde{f}(p) &= a^{4}\sum_{x \in \Lambda_{a}} e^{ip\cdot x}f(x),\\
f(x) &= \frac{1}{a^{4}N_{s}^{3}N_{\tau}}\sum_{p \in \mathcal{B}_{a}} e^{-ip\cdot x}\widetilde{f}(p),
\label{eq:fourier}
\end{align}
with momenta restricted to the first Brillouin zone $\mathcal{B}_{a} = \left\{p \in \mathbb{R}^{4} : -\frac{\pi}{a} < p_{\mu} \leq \frac{\pi}{a} \right\}$. A finite lattice
spacing thus implies a momentum cutoff $\Lambda=\pi/a$.\\

\noindent
While there is no rigorous proof to date, there is plenty of evidence from different computational techniques that the continuum limit of four-dimensional $\phi^{4}$ theory is trivial, i.e, it corresponds to a free theory with renormalised coupling $g_{R}=0$. If the momentum cutoff is chosen finite but high enough, corresponding to a small fixed lattice spacing $a$, it is possible to have an interacting theory with practically negligible discretisation effects. The theory then represents a non-trivial effective field theory with $g_{R} \neq 0$ up to that cutoff $\Lambda$~\cite{Montvay:1987us,Montvay:1994cy}. There are then three relevant scales in the problem: the physical mass of the scalar particle, $m$, the ultraviolet cutoff fixed by the lattice spacing, $\Lambda=\pi/a$, and the temperature, $T$. Since our aim is to study the general properties of perturbation theory rather than a specific physics application, we refrain from setting an absolute scale and express all numerical results as dimensionless ratios. \\

\noindent
Our observable of interest is the Euclidean two-point correlation function $\langle \phi(\tau,\vec{x})\phi(0)\rangle$, which carries information about the thermal medium and is easily accessible by both perturbation theory and numerical simulations. For the following we recall that on a Euclidean space-time lattice the thermal partition function $Z$ can be represented equivalently by two different Hamiltonians,
\begin{align}
Z=\mathrm{Tr}(e^{-aH N_{\tau}})=\mathrm{Tr}(e^{-aH_{z} N_{z}}).
\end{align}
$H$ is the usual Hamiltonian translating states by one lattice spacing in Euclidean time, whereas $H_{z}$ is the analogous operator translating states in the $z$ direction,
\begin{align}
&|\psi(\tau+a, x,y,z)\rangle = e^{-aH}|\psi(\tau,x,y,z)\rangle, \\
&|\psi(\tau, x,y,z+a)\rangle = e^{-aH_{z}}|\psi(\tau,x,y,z)\rangle. 
\end{align}
Both are straightforwardly defined via the transfer matrix between adjacent $\tau$ or $z$ slices, respectively~\cite{Montvay:1994cy}, and are related by relabelling $\tau\leftrightarrow z$. The thermodynamic limit ($N_{x,y,z} \rightarrow \infty$, with $T^{-1}=aN_\tau$ finite) formally represents the ``vacuum'' physics of $H_{z}$, whose spectrum is sensitive to the ``finite volume'' effect of the compactified $\tau$ direction. Screening masses are the corresponding ground state energies of $H_{z}$ in each quantum number channel, and in the limit $T \rightarrow 0$ their spectrum is identical to the mass spectrum of $H$. Inserting a complete set of eigenstates of $H_{z}$, $H_{z}|n\rangle = E_{n}|n\rangle$, the spectral representation of the two-point function at $(\tau,\vec{x})=(0,\vec{z})=(0,0,0,z)$ then reads
\begin{align}
\langle \phi(0,\vec{z})\phi(0)\rangle = \frac{\sum_{m,n}|\langle m|\phi(0)|n\rangle|^2 e^{-(E_{n}-E_{0})z} e^{-(E_{m}-E_{0})(L-z)}}{1+e^{-L(E_{1}-E_{0})}+\ldots} .
\end{align}
In the vacuum and infinite volume limits the energy eigenvalues are identical to those of the true Hamiltonian on account of Euclidean rotation symmetry. In particular, the first energy level above the vacuum is the mass gap $m=E_{1}-E_{0}$, i.e.~the physical mass of the scalar particle, its inverse is the correlation length of the system. When $N_{s}$ is finite there are explicit and implicit finite size corrections $E_{n}(L)$, both of which are exponentially small once $N_{s}$ exceeds a few times the correlation length, $mL\sim \mathcal{O}(1)$~\cite{Montvay:1987us,Montvay:1994cy}. On the other hand, finite $N_{\tau}$ corresponds to a finite temperature, whose effects are implicit in the eigenvalues of $H_{z}$. These eigenvalues $E_{n}=E_{n}(T)$ no longer describe the vacuum spectrum, but those of the inverse spatial screening lengths in a medium. The inverse correlation length of the finite temperature system is given by the lowest screening mass, $m(T)=E_{1}(T)-E_{0}(T)$, which in the absence of phase transitions smoothly approaches the particle mass in the limit $T\rightarrow 0$. Although in numerical simulations both $N_{\tau}$ and $N_{s}$ are necessarily finite, the infinite $N_{\tau}$ and $N_{s}$ correlators can be approximated arbitrarily well up to exponentially small corrections. Technical details of these simulations are discussed in Appendix~\ref{lattice_details}. \\

\noindent
Our main focus in this work is the infrared structure of finite-temperature QFT, and for this purpose we focus on the \textit{spatial} correlator, where the lattice two-point function is summed over both temporal and orthogonal spatial directions   
\begin{align}
C(z;a,N_{s},N_{\tau}) &=  a^{3}\sum_{\tau,x,y}\langle \phi(\tau,\vec{x})\phi(0)\rangle.
\end{align}  
This corresponds to a partial Fourier transform projecting onto the zero-momentum ($\omega_{E}=p_{x}=p_{y}=0)$ states of $H_{z}$.

\subsection{Lattice perturbation theory at zero and finite $T$}
\label{sec:pt}

In order to cleanly separate the convergence properties of perturbation theory from any other systematics, we work with lattice perturbation theory for a finite box specified by $N_{s}$ and $N_{\tau}$. In this case there are no finite volume or discretisation effects between the perturbative and simulated results. Moreover, this is the most general situation, since the infinite volume and vacuum limits arise as special cases. This implies in particular that there is no technical difference between Euclidean directions: spatial momenta are discretised to Matsubara-like frequencies just as the temporal components are, viz. Eq.~\eqref{eq:fourier}, and so the computational scheme is precisely the same for hot or cold, and small or large systems. By employing the same lattice regularisation, perturbative and numerical predictions for identical bare parameter sets can therefore be directly compared, and hence any statistically significant difference must be entirely due to the perturbative approximation. \\

\begin{figure}[t!] 
\centering
\includegraphics[width=0.5\textwidth]{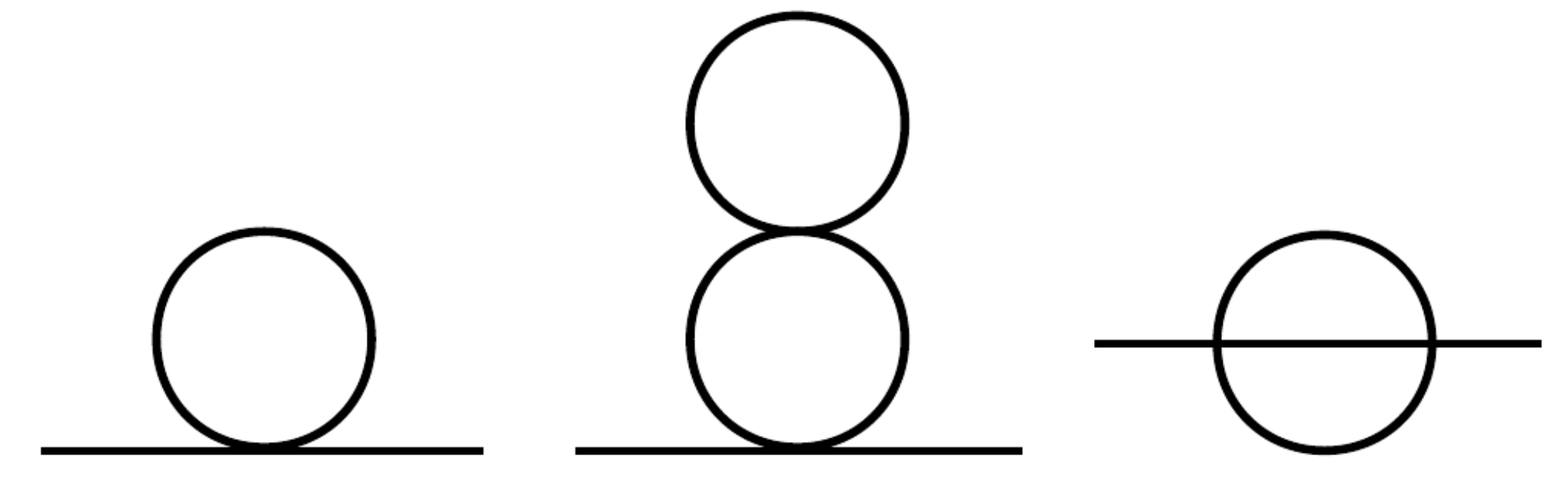}
\caption{The contributing diagrams to the renormalised self-energy up to $\mathcal{O}(g_{0}^{2})$. From left to right these are the tadpole, cactus, and sunset diagrams, respectively.}
\label{self_energy_diag}
\end{figure}

\noindent
On the lattice the free-field Feynman propagator takes the form
\begin{align}
\widetilde{G}_{0}(p;a,N_{s},N_{\tau}) = \frac{1}{\sum_{\mu}\tfrac{4}{a^{2}}\sin^{2}\left(\tfrac{a p_{\mu}}{2}\right)+m_{0}^{2}} \ \stackrel{a\rightarrow 0}{\longrightarrow} \
\frac{1}{\omega_{E}^{2}+ |\vec{p}|^{2}+m_0^{2}},
\end{align}
which approaches the standard continuum Euclidean propagator in the limit of vanishing lattice spacing. In the literature, mostly the massless case $m_0=0$ is discussed,
for which the zero mode $p=0$ is divergent, motivating all-order resummations. Here we keep $m_0>0$ throughout, thus avoiding this issue. This also means that there is no parametric mixing (in powers of the coupling) of different loop orders for soft momenta, and hence any resummation is unnecessary. Defining the self energy $\Pi(p;a,N_{s},N_{\tau})$ as the sum of all one-particle irreducible diagrams, the full propagator of the interacting theory corresponds to a geometrical series in $\Pi$, as in the continuum, which can be summed to
\begin{align}
\widetilde{G}(p;a,L,N_{\tau}) =  \frac{1}{\sum_{\mu}\tfrac{4}{a^{2}}\sin^{2}\left(\tfrac{a p_{\mu}}{2}\right)+m_{0}^{2} + \Pi(p;a,N_{s},N_{\tau}) }.
\label{prop_full}
\end{align}
In general, the self-energy representation in Eq.~\eqref{prop_full} can be used to write the spatial correlator in the following form
\begin{align}
C(z;a,N_{s},N_{\tau}) &=  \frac{1}{N_{s}} \! \sum_{k_{z}=0}^{N_{s}-1} \! e^{\frac{2\pi i k_{z}}{aN_{s}}z}  \frac{a}{4\sin^{2}\!\left(\tfrac{\pi k_{z} }{N_{s}}\right)+(am_{0})^{2}+a^{2}\Pi(\omega_{E}=p_{x}=p_{y}=0,p_{z}=\frac{2\pi k_{z}}{aN_{s}};a,N_{s},N_{\tau})}.  
\label{Cz}
\end{align}
By evaluating $\Pi$ up to some fixed order in the bare coupling $g_{0}$, Eq.~\eqref{Cz} can then be used to calculate the spatial correlator up to this order. Up to $\mathcal{O}(g_{0}^{2})$, viz.~two-loop order, the contributing diagrams are the one-loop tadpole, and the two-loop cactus and sunset diagrams, as displayed in Fig.~\ref{self_energy_diag}. 
The two-loop self-energy then takes the explicit form~\cite{Montvay:1994cy} 
\begin{align}
a^{2}\Pi_{(2)}(p;a,N_{s},N_{\tau}) &= \frac{g_{0}}{2}J_{1}(am_{0};N_{s},N_{\tau}) -\frac{g_{0}^{2}}{4}J_{1}(am_{0}; N_{s},N_{\tau})J_{2}(am_{0}; N_{s},N_{\tau})  -\frac{g_{0}^{2}}{6}I_{3}(p,am_{0}; N_{s},N_{\tau}),
\label{2loop_SE}
\end{align}
where the tadpole, cactus, and sunset contributions correspond to the first, second, and third terms, respectively. The real-valued functions $J_{n}$ and $I_{3}$ are defined as
\begin{align}
J_{n}(am_{0};N_{s},N_{\tau}) &= \frac{1}{N_{s}^{3}N_{\tau}}\sum_{p \in \mathcal{B}_{a}} \frac{1}{\left[\sum_{\mu} 4\sin^{2}\left(\tfrac{a p_{\mu}}{2}\right)+(am_{0})^{2}\right]^{n}}, \quad\quad n=1,2 \label{Jn} \\
I_{3}(p,am_{0};N_{s},N_{\tau}) &= \frac{1}{(N_{s}^{3}N_{\tau})^{2}} \sum_{q\in \mathcal{B}_{a}}  \sum_{r\in \mathcal{B}_{a}}   \frac{1}{\left[\sum_{\mu} 4\sin^{2}\left( p_{\mu}-\tfrac{a q_{\mu}}{2}-\tfrac{a r_{\mu}}{2}\right)+(am_{0})^{2}\right]}  \nonumber \\
& \quad\quad\quad\quad\quad\quad \times \frac{1}{\left[\sum_{\nu} 4\sin^{2}\left(\tfrac{a q_{\nu}}{2}\right)+(am_{0})^{2}\right]}   \frac{1}{\left[\sum_{\Lambda_{a}} 4\sin^{2}\left(\tfrac{a r_{\lambda}}{2}\right)+(am_{0})^{2}\right]}. \label{I3}
\end{align}

\ \\

\begin{figure}[t!] 
\centering
\includegraphics[width=0.4\textwidth]{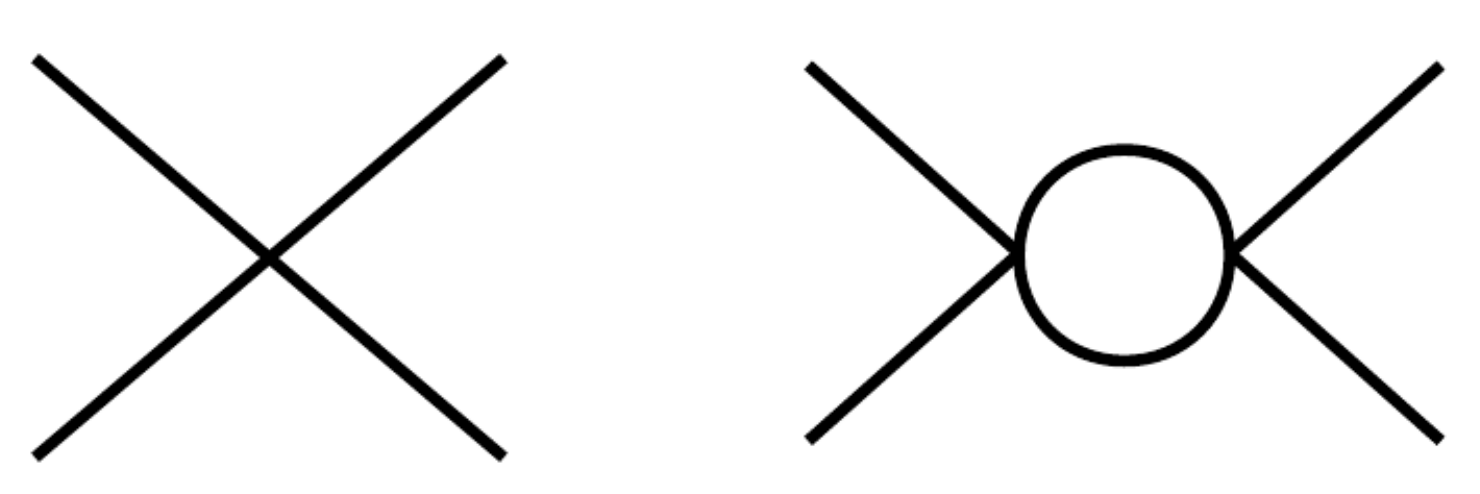}
\caption{The contributing diagrams to the renormalised coupling $g_{R}$ up to $\mathcal{O}(g_{0}^{2})$.}
\label{coupling_diag}
\end{figure}

\noindent
In order to quantify the physical coupling strength of the theory, we define the renormalised coupling through the four-point function evaluated at zero momentum, whose explicit expression for finite $N_{s}$ and $N_{\tau}$ up to order $\mathcal{O}(g_{0}^{2})$ reads~\cite{Montvay:1994cy} 
\begin{align}
g_{R}\equiv -\Gamma^{(4)}_{R}(0,0,0,0) = g_{0}-\frac{3}{2}g_{0}^{2} J_{2}(am_{0};N_{s},N_{\tau}) + \mathcal{O}(g_{0}^{3}),
\label{eq:g_R}    
\end{align}
the contributions of which correspond to the perturbative diagrams in Fig.~\ref{coupling_diag}. \\

\noindent
The exponential decay of the two-point function in coordinate space is governed by its pole mass in momentum space, which represents the particle mass in vacuum. Particularly interesting in the current context is the one-loop result $m_{(1)}$ which arises from the tadpole diagram~\cite{Montvay:1994cy}:
\begin{align}
m_{(1)}=\frac{2}{a}\ln \left(\frac{a m_{R}}{2}+\sqrt{1+\frac{(a m_{R})^2}{4}}\right) + \mathcal{O}(g_{R}^{2}),\quad  m_{R}^{2} = m_{0}^{2}+\frac{g_{0}}{2a^{2}}J_{1}(am_{0};N_{s},N_{\tau}) + \mathcal{O}(g_{0}^{2}).
\end{align}
As in the continuum, the correlator has a real pole at this level, corresponding to a shift in the vacuum mass from the momentum-independent tadpole contribution. That the renormalised one-loop self-energy $\Pi_{(1)}$ is constant is another indication besides the NRT theorem that the standard finite-temperature perturbative approach is inconsistent. This is because the derivation of the perturbative expansion follows from the Gell-Mann-Low relation~\cite{Gell-Mann:1951ooy}, which relies on the assumption of asymptotic on-shell scattering states. If these states exist, then the inverse propagator must vanish on the mass shell of these states at each perturbative loop order, which is not true here since $\Pi_{(1)}$ is non-vanishing and momentum-independent~\cite{Landsman:1988ta}.
    
\section{Perturbation theory vs.~numerical simulations}
\label{lattice_analysis}

The predictive power of any given order of a weak coupling expansion depends on the interaction strength of the theory and is smoothly connected to the free-field result in the limit of vanishing interactions. In this section we quantify the success of perturbation theory by the deviation of its prediction at fixed order in the coupling from the exact answer, which is known up to a statistical error from numerical simulations. To this end we evaluate the relative deviation of the perturbative $n$-loop spatial correlator $C^{(n)}(z)$ from the Monte-Carlo simulated correlator $C^{\text{sim}}(z)$ at a correlation distance of $zm=1$, where $m$ is the screening mass which is determined by fitting an exponential form to the spatial correlator data for sufficiently large $N_{\tau}$ such that $m$ no longer deviates significantly. The scale $m$ serves as a good approximation to the physical mass in the true vacuum limit $N_{\tau}\rightarrow \infty$. The relative deviation at the point $zm=1$ is defined 
\begin{align}
\Delta_{n} = \frac{C^{(n)}(zm=1)-C^{\text{sim}}(zm=1)}{C^{\text{sim}}(zm=1)}.
\end{align}
The choice $zm=1$ on the one hand represents a distance short enough to still have high statistical accuracy in the Monte-Carlo correlator, while on the other hand it corresponds to one Compton wavelength of the vacuum particle, or one screening length at low temperature, and hence marks the onset of the thermally interesting long-distance physics. Since our $z$-variable is discrete, we take the smallest lattice distance greater than one Compton wavelength. As a second quality measure, we evaluate the $\chi^{2}/\text{d.o.f.}$ with which the two-loop correlator coincides with the simulated data over the entire available distance range.

\subsection{Vacuum predictions}

Before investigating temperature effects we establish the vacuum situation as a baseline for our comparison. We begin by illustrating the gradually changing quality of perturbation theory by considering different renormalised coupling strengths $g_{R}$, for which we use the $\mathcal{O}(g_{0}^{2})$ perturbative result in Eq.~\eqref{eq:g_R}, while keeping other physical parameters fixed. For this purpose we consider the three bare parameter sets specified in Table~\ref{tab:vacL16}, corresponding to nearly constant mass scales $m/\Lambda$ with varying values of $g_{R}$, all on the same $16^{3} \times 16$ lattice. The UV cutoff for the theory $\Lambda$ is nearly a factor of ten above the scalar particle mass $m$, and should therefore have negligible influence on the low-energy physics. Furthermore, finite volume and temperature effects are exponentially suppressed since $mL,m/T>5$. \\

\begin{table}[h!]
\center
\small
\renewcommand{\arraystretch}{1.15}
\begin{tabular}{|c|c|c|c|c|c|c|c|c|} 
\hline
\rule{0pt}{3ex}
$(am_{0}, g_{0})$ &  $g_{R}$ & $m/\Lambda$ & $mL$  & $T/m$ & $\left(\chi^{2}/\text{d.o.f.}\right)_{\text{2-loop}}$ & $\Delta_{0}$ [\%] & $\Delta_{1}$ [\%] &  $\Delta_{2}$ [\%]  \\[0.5ex]
\hhline{|=|=|=|=|=|=|=|=|=|}
(0.315, 0.5) & 0.49 & 0.118  & 5.93   &  0.17   &  1.7  &   42.3 (0.9)  & 0.4 (0.6)   & 0.8 (0.6)   \\ 
\hline
(0.25, 1.0)  & 0.94 & 0.117  & 5.88  & 0.17    &  0.1  &  127.0 (1.4)  & 2.0 (0.6)   & 0.2 (0.6)   \\ 
\hline 
(0.15, 1.5)  & 1.25 & 0.115  & 5.78  & 0.17      &  10.3 &  509.8 (3.5)  & 4.1 (0.6)   & 2.3 (0.6)   \\
\hline
\end{tabular}
\caption{The reduced $\chi^{2}$ values of the two-loop perturbative prediction and relative deviations of the leading order, $\Delta_{0}$, one-loop, $\Delta_{1}$, and two-loop, $\Delta_{2}$, predictions at a correlation distance of $zm=1$. The numbers in brackets indicate the uncertainty on these relative deviations. The uncertainties on $m/\Lambda$, $mL$, and $T/m$ are not displayed because they are smaller than the accuracy shown in the table.}
\label{tab:vacL16}
\end{table} 

\begin{figure}[t!]
\centering
\includegraphics[width=0.45\textwidth]{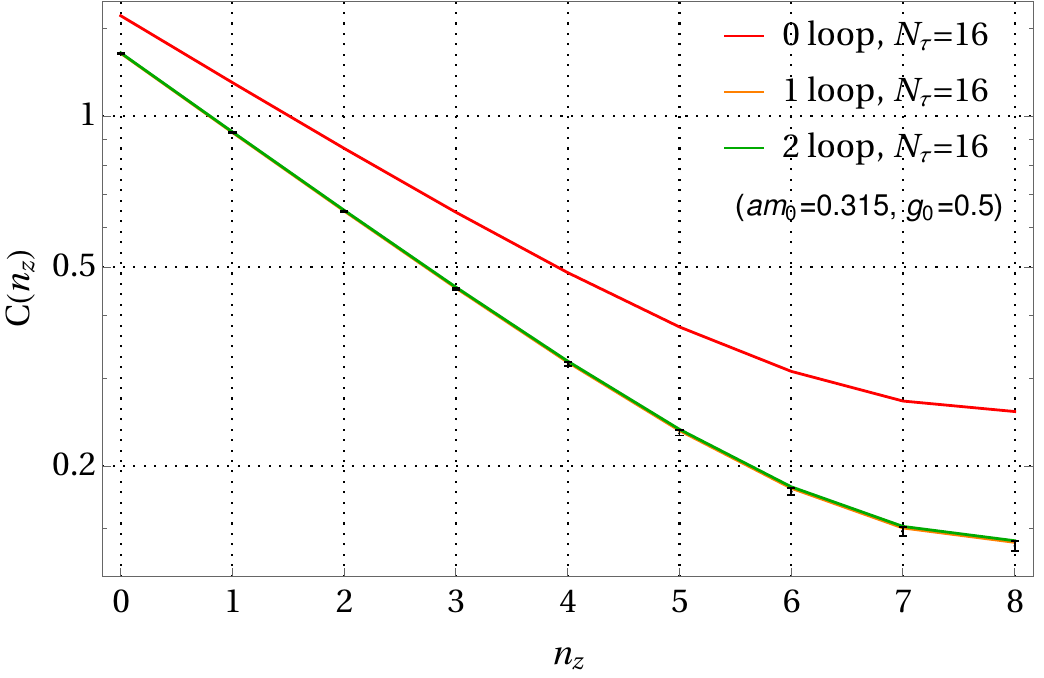}
\includegraphics[width=0.45\textwidth]{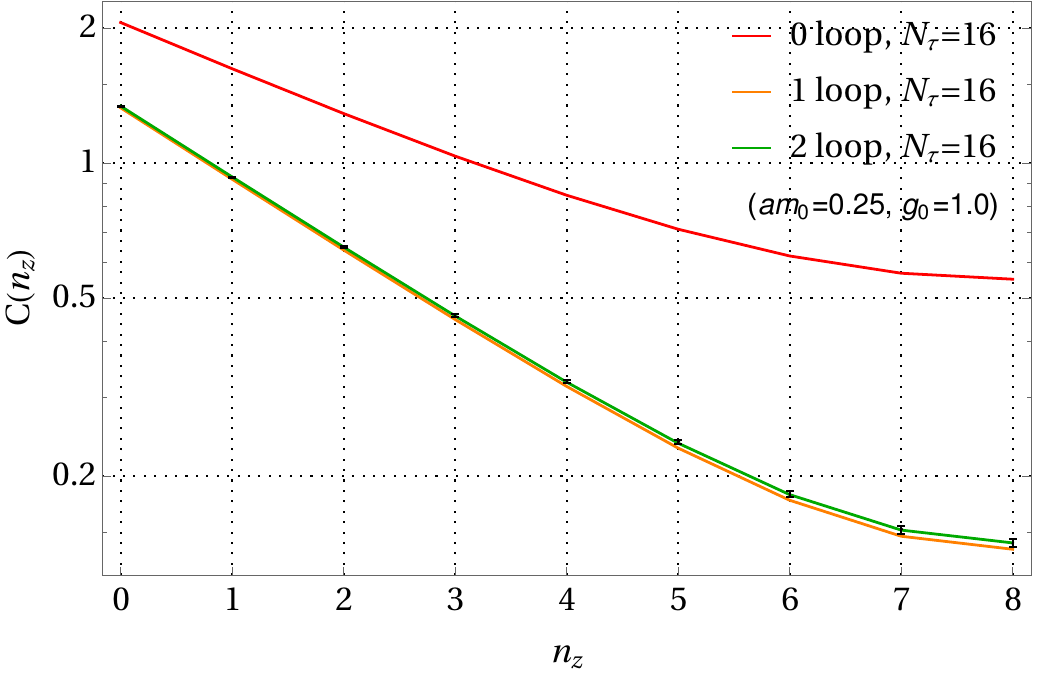}
\includegraphics[width=0.45\textwidth]{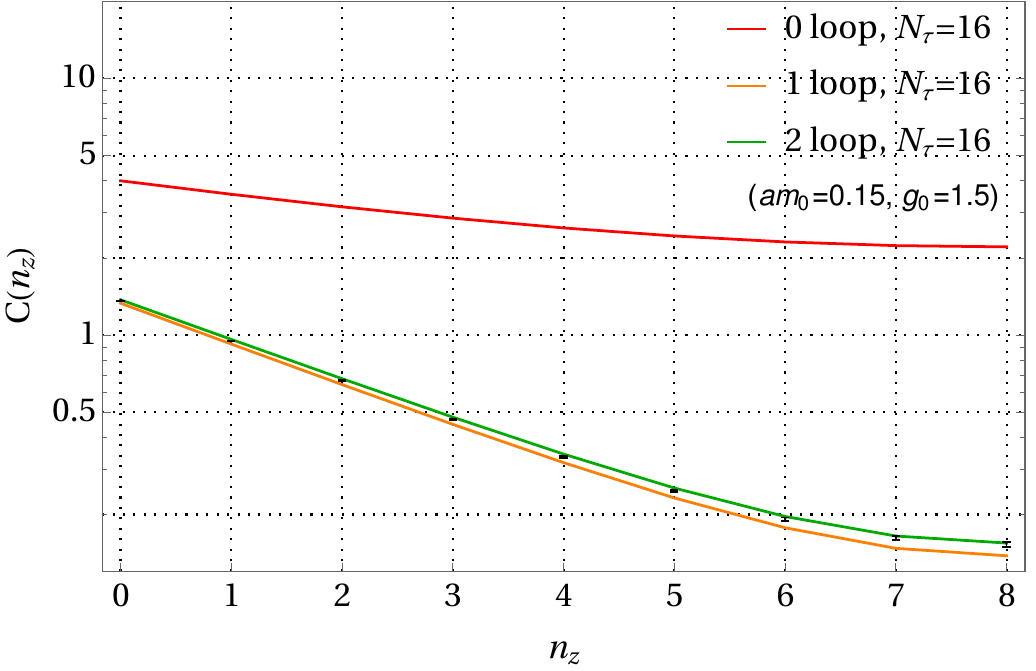}
\caption{Comparison of the perturbative $C(n_{z})$ predictions with the different $16^3\times 16$ lattice data simulations (black points).}
\label{fig:vac16}
\end{figure}

\noindent
In Fig.~\ref{fig:vac16} the perturbative correlator predictions are compared with the simulated correlators using the different parameter sets in Table~\ref{tab:vacL16}. The perturbative predictions at each discrete $n_{z}$ value are joined together in the figures for ease of comparison. From this figure one can clearly see that the coupling strength plays an important role in all three parameter sets. At the strongest coupling (bottom plot) the series displays an asymptotic-like behaviour, with the one-loop and two-loop corrections coming with different signs, but getting rapidly smaller in magnitude. As the coupling gets weaker the differences between one and two-loop contributions are no longer distinguishable by eye. This is more clearly elucidated in Table~\ref{tab:vacL16}: at the given numerical precision the first parameter set reproduces the simulated correlator within errors at one-loop level already, for the second set this is only the case at two-loop level, whilst for the strongest coupling the two-loop result is outside of the error bars, leading to the large reduced $\chi^{2}$. It should be noted though that even in this case the relative deviation at one screening length is still only $\approx 2\%$. In other words, for all three parameter sets perturbation theory works well and is quantitatively accurate.

\subsection{Finite temperature predictions}
\label{sec:fT}

\begin{table}[t!]
\center
\small
\renewcommand{\arraystretch}{1.16}
\begin{tabular}{|c|c|c|c|c|c|c|c|c|} 
\hline
\rule{0pt}{3ex}
$N_{s}^{3} \times N_{\tau}$ &  $(am_{0}, g_{0})$ & $g_{R}$ &  $m/\Lambda$  & $mL$  & $T/m$  & $\left(\chi^{2}/\text{d.o.f.}\right)_{\text{2-loop}}$ & $\Delta_{1}$ [\%] & $\Delta_{2}$ [\%]  \\[0.5ex]
\hhline{|=|=|=|=|=|=|=|=|=|}
$16^{3} \times 8$      & $(0.315, 0.5)$  & 0.49 &  0.118  & 5.93	&     0.34   &  0.4    & 0.2 (0.6)   & 0.2 (0.6)   \\
\hline
$16^{3} \times 4$      & $(0.315, 0.5)$  & 0.49 &  0.118  & 5.93	&     0.68   &  0.3    & 1.0 (0.6)   & 0.4 (0.6)   \\
\hline
$16^{3} \times 2$      & $(0.315, 0.5)$  & 0.49 &  0.118  & 5.93    &     1.35   &  0.5    & 1.2 (0.6)   & 0.4 (0.6)   \\ 
\hhline{|=|=|=|=|=|=|=|=|=|}
$16^{3} \times 8$      & $(0.25, 1.0)$   & 0.94 &  0.117  & 5.88    &     0.34   &  0.7    & 1.1 (0.6)   & 0.8 (0.6)  \\
\hline
$16^{3} \times 4$      & $(0.25, 1.0)$   & 0.94 &  0.117  & 5.88    &     0.68   &  0.1    & 2.4 (0.6)   & 0.1 (0.6)   \\
\hline
$16^{3} \times 2$      & $(0.25, 1.0)$   & 0.94 &  0.117  & 5.88    &     1.36   &  3.9    & 3.3 (0.6)   & 0.9 (0.6)  \\
\hhline{|=|=|=|=|=|=|=|=|=|}
$16^{3} \times 8$      & $(0.15, 1.5)$   & 1.25 &  0.115  & 5.78    &     0.35   &  49.7   & 4.6 (0.6)   & 4.2 (0.6)   \\
\hline
$16^{3} \times 4$      & $(0.15, 1.5)$   & 1.25 &  0.115  & 5.78    &     0.69   &  188.5  & 6.0 (0.6)   & 8.7 (0.7) \\
\hline 
$16^{3} \times 2$      & $(0.15, 1.5)$   & 1.25 &  0.115  & 5.78    &     1.38   &  1262.3 & 8.7 (0.5)   & 22.2 (0.7)  \\ 
\hline
\end{tabular}
\caption{The reduced $\chi^{2}$ values of the two-loop perturbative prediction and relative deviations of the one $\Delta_{1}$ and two-loop $\Delta_{2}$ predictions at a correlation distance of $zm=1$. The numbers in brackets indicate the uncertainty on these relative deviations. The uncertainties on $m/\Lambda$, $mL$, and $T/m$ are not displayed because they are smaller than the accuracy shown in the table.}
\label{tab:TL16}
\end{table} 

\begin{figure}[t!]
\centering
\includegraphics[width=0.45\textwidth]{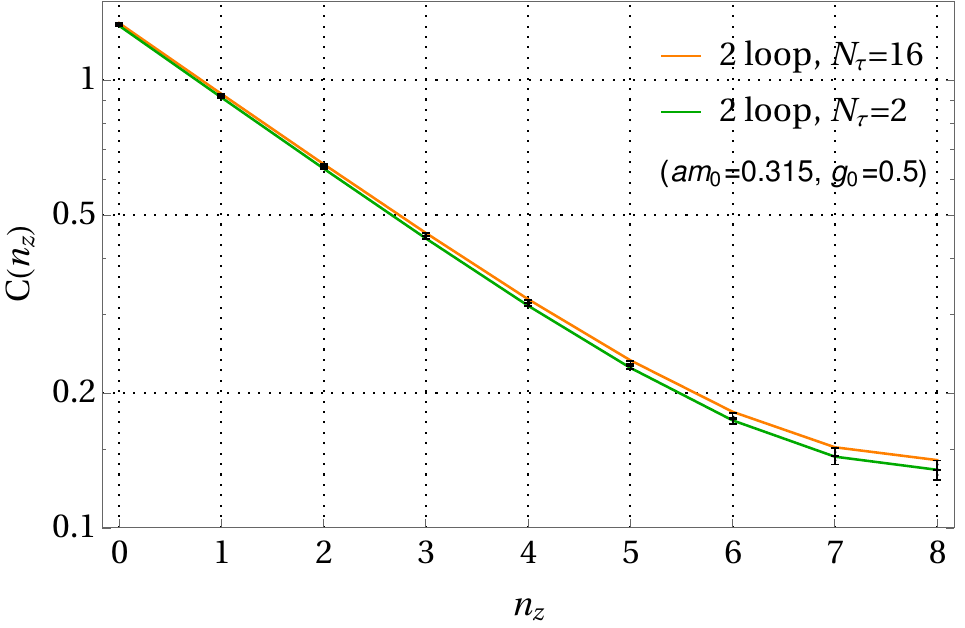}
\includegraphics[width=0.45\textwidth]{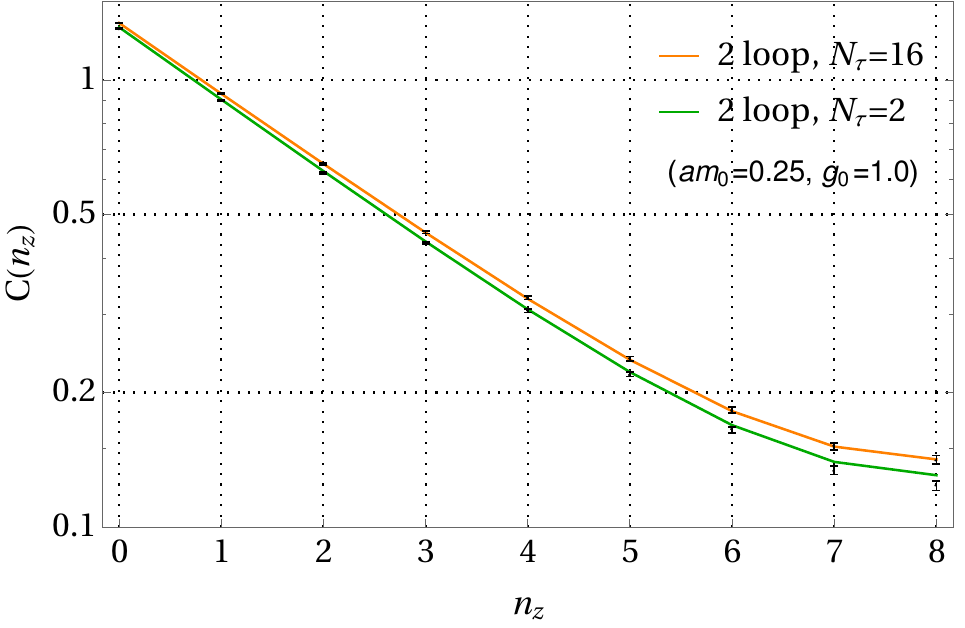}
\includegraphics[width=0.45\textwidth]{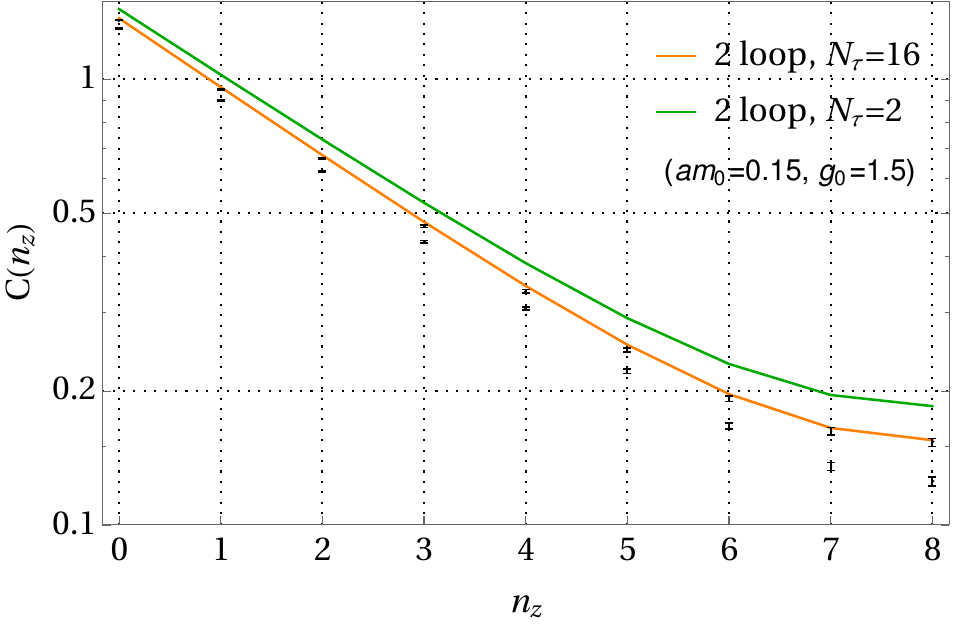}
\caption{Comparison of the two-loop perturbative $C(n_{z})$ predictions with the different $N_{s}=16$ lattice simulations (black points). The upper data points ($N_{\tau}=16$) agree well for all bare parameter sets, whereas the lower points ($N_{\tau}=2$) display deviations from the two-loop predictions, and these are particularly significant for the parameter set in the lower panel.}
\label{fig:L16T16T2}
\end{figure}

To explore the impact of temperature we now gradually reduce $N_{\tau}$ while keeping all other parameters fixed. The results are displayed in Table~\ref{tab:TL16} and plotted in Fig.~\ref{fig:L16T16T2} for the largest and smallest $N_{\tau}$ values. This corresponds to the so-called fixed-scale approach, where the lattice spacing remains constant and temperature can only be changed in discrete steps. Note that the temperatures which can be achieved in this way are limited to the regime $0<T/m \lesssim \mathcal{O}(1)$. Nevertheless, these results show an unambiguous trend: \textit{the perturbative predictions increasingly deteriorate as the temperature is increased.} This is most clearly visible in the values of the reduced $\chi^{2}$ and relative deviations at one Compton wavelength. The only parameter set without significant deviations is the one with the weakest coupling, where $g_{R}=0.49$. However, as the corresponding figure shows (top left in Fig.~\ref{fig:L16T16T2}), at such a weak coupling there is barely any temperature effect up to $T/m=1.35$, and the correlator at this temperature only slightly differs from the vacuum-like one. For each of the other couplings the relative deviation of the one-loop and two-loop predictions have grown by an order of magnitude at the same temperature. Moreover, for the largest coupling, the prediction at two-loop order is actually worse than at one loop, and so the asymptotic-like behaviour displayed by the vacuum series appears to be lost. This is a reflection of the two-loop prediction giving the wrong qualitative behaviour, namely that the finite-temperature correlator is above the vacuum one, contrary to the data.  \\

\begin{figure}[t!]
\includegraphics[width=0.47\textwidth]{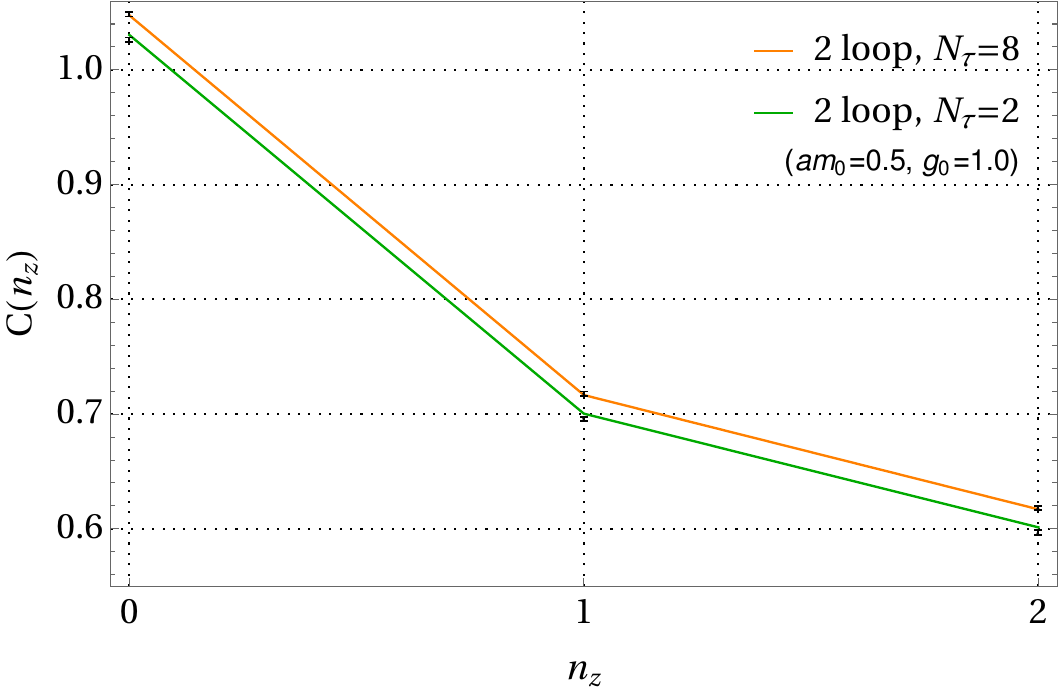}
\includegraphics[width=0.49\textwidth]{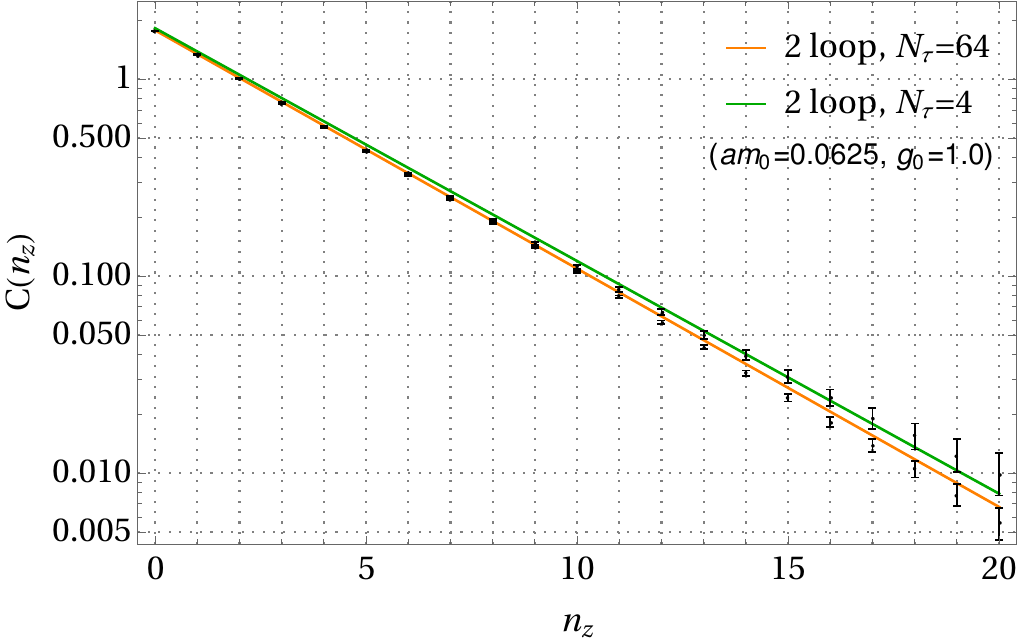}
\caption{Comparison of the two-loop perturbative $C(n_{z})$ predictions with the lattice data simulations for $N_{s}=4$ (left panel) and $N_{s}=64$ (right panel). The upper and lower black points indicate the spatial correlator data for the largest and smallest temporal sizes, respectively.}
\label{fig:L64T64T4}
\end{figure}

\noindent
That these findings are related to temperature alone, and independent of the spatial size of the system, is demonstrated by two further parameter sets for which a similar series of slowly increasing temperatures were investigated, cf.~Table~\ref{tab:TL64:TL4} and Fig.~\ref{fig:L64T64T4}. Note that in these sets the spatial box sizes are vastly different from those in the previous cases, with $mL\sim 2$ and $mL\sim18$, respectively. For the vacuum-like lattices, perturbation theory consistently captures the finite-size effects, see also~\cite{Montvay:1987us,Montvay:1994cy}. But as in the previous cases, as soon as $T/m\sim \mathcal{O}(1)$ the asymptotic-like convergence behaviour is lost. This is particularly apparent for the $N_{s}=64$ lattice at the highest temperature, where the two-loop prediction once again deviates further from the data than that at one loop, and the correlator displays the wrong ordering relative to the vacuum result.   \\

\begin{table}[t!]
\center
\small
\renewcommand{\arraystretch}{1.16}
\begin{tabular}{|c|c|c|c|c|c|c|c|c|} 
\hline
\rule{0pt}{3ex}
$N_{s}^{3} \times N_{\tau}$ &  $(am_{0}, g_{0})$ & $g_{R}$ & $m/\Lambda$  & $mL$  & $T/m$  & $\left(\chi^{2}/\text{d.o.f.}\right)_{\text{2-loop}}$ & $\Delta_{1}$ [\%] &  $\Delta_{2}$ [\%]  \\[0.5ex]
\hhline{|=|=|=|=|=|=|=|=|=|}
$4^{3} \times 8$       & $(0.5, 1.0)$    &  0.91  &  0.178   &  2.24  &  0.22   &  0.5    & 1.3 (0.3)   & 0.2 (0.3)  \\
\hline
$4^{3} \times 4$       & $(0.5, 1.0)$    &  0.91  &  0.178   &  2.24  &  0.45   &  0.4    & 1.5 (0.3)   & 0.1 (0.3) \\
\hline
$4^{3} \times 2$       & $(0.5, 1.0)$    &  0.91  &  0.178   &  2.24  &  0.89   &  7.7    & 2.3 (0.3)   & 0.8 (0.3)  \\
\hhline{|=|=|=|=|=|=|=|=|=|}
$64^{3} \times 64$     & $(0.0625, 1.0)$ &  0.91  &  0.088   & 17.77  &  0.06   &  2.0    & 3.4 (0.5)  & 0.6 (0.5)  \\ 
\hline
$64^{3} \times 32$     & $(0.0625, 1.0)$ &  0.91  &  0.088   & 17.77  &  0.11   &  3.4    & 3.3 (0.3)  & 0.9 (0.4) \\
\hline 
$64^{3} \times 16$     & $(0.0625, 1.0)$ &  0.91  &  0.088   & 17.77  &  0.23   &  24.2   & 3.3 (0.2)  & 1.8 (0.3) \\
\hline
$64^{3} \times 8$      & $(0.0625, 1.0)$ &  0.91  &  0.088   & 17.77  &  0.45   &  168.2  & 3.8 (0.2)  & 3.3 (0.2) \\
 \hline
$64^{3} \times 4$      & $(0.0625, 1.0)$ &  0.91  &  0.088   & 17.77  &  0.90   &  588.9  & 4.6 (0.2)  & 7.6 (0.2) \\
\hline
\end{tabular}
\caption{The reduced $\chi^{2}$ values of the two-loop perturbative prediction and relative deviations of the one $\Delta_{1}$ and two-loop $\Delta_{2}$ predictions at a correlation distance of $zm=1$. The numbers in brackets indicate the uncertainty on these relative deviations. The uncertainties on $m/\Lambda$, $mL$, and $T/m$ are not displayed because they are smaller than the accuracy shown in the table.}
\label{tab:TL64:TL4}
\end{table}

\noindent
In summary, by comparing the predictions of perturbation theory with lattice simulations, both in vacuum and at finite temperature, we observe the following distinct characteristics:

\begin{enumerate}
\item For sufficiently small coupling values the perturbative predictions are consistent with the data for all temperatures. This is not surprising, since the contribution of interaction terms in Eq.~\eqref{2loop_SE} become increasingly negligible as $g_{0}\rightarrow 0$, and hence the system approaches that of a free particle, which experiences no thermal modifications. The smallness of the coupling is also reflected in the comparable accuracy of the one and two-loop results.

\item For moderate coupling values the two-loop predictions begin to deviate beyond the errors of the vacuum data, although the two-loop results are more accurate than those at one loop, and higher-order corrections would be expected to further increase the precision of these predictions.

\item In all parameter sets where the effects of interactions are non-negligible the perturbative predictions increasingly deteriorate as the temperature of the system increases. This feature is most pronounced in the final parameter sets of Tables~\ref{tab:TL16} and~\ref{tab:TL64:TL4}, which are plotted in Figs.~\ref{fig:L16T16T2} and~\ref{fig:L64T64T4}. Here the relative deviations of the two-loop predictions at the highest temperatures are actually \textit{larger} than those at one loop, and hence higher-order corrections are driving the predictions further away from the data such that even the temperature ordering of the correlators is incorrect. This reflects a \textit{qualitative} change from the behaviour of the series in vacuum, and a breakdown of the perturbative treatment.

\end{enumerate}

\subsection{Origin of the problem}
\label{origin}

\begin{figure}[h!]
\centering
\includegraphics[width=0.6\textwidth]{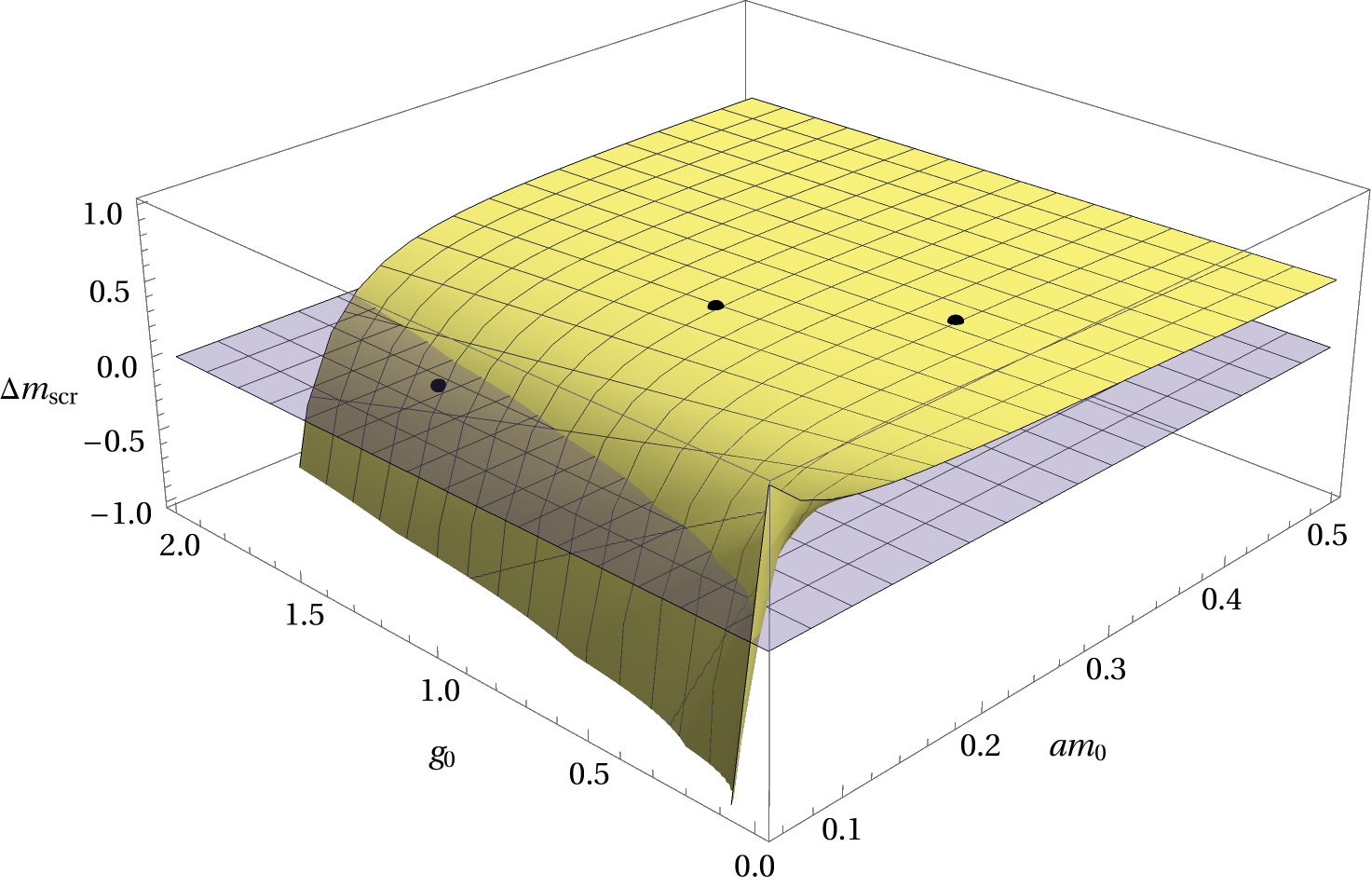}
\caption{Plot of $\Delta m_{\text{scr}}$ for $N_{s}=16$ as a function of the lattice bare parameters $(am_{0},g_{0})$. The intersection of $\Delta m_{\text{scr}}$ with the $\Delta m_{\text{scr}}=0$ plane (light blue) indicates the boundary of the region $\Delta m_{\text{scr}}<0$, and the black points are the three bare parameter sets used.}
\label{amg0Screen}
\end{figure}

In the previous section we found that the two-loop perturbative predictions provided a consistent description of the lattice data for vacuum-like systems at moderate coupling values, but as the temperature increased these predictions deteriorated. A strong indication of the theoretical origin of these deviations can be seen in the $N_{s}=16$ case, where the physical values of $T/m$ between the different parameter sets are comparable, and yet the deviations from the numerical results are significantly different depending on $N_{\tau}$. Since the spatial correlator can be written in the form of Eq.~\eqref{Cz}, these deviations must ultimately be driven by the properties of the self-energy $\Pi(p;a,N_{s},N_{\tau})$, which at two-loop order is fixed by the functions $J_{n}(am_{0};N_{s},N_{\tau})$ and $I_{3}(p,am_{0};N_{s},N_{\tau})$. In each of these cases it turns out that the contribution of the two-loop sunset diagram in Eq.~\eqref{2loop_SE} is relatively small, and that the deviations in the two-loop predictions are driven by the interplay between the tadpole and cactus diagrams. For a certain choice of bare lattice parameter values, the contribution of these diagrams to $\Pi(p;a,N_{s},N_{\tau})$ at each fixed value of $N_{\tau}$ is \textit{less} than the contribution in the zero-temperature $N_{\tau}\rightarrow \infty$ limit. Neglecting the sunset diagram, this implies that the screening mass $m_{\text{scr}}$, which in the continuum controls the exponential-like behaviour of the spatial correlator $C(z) \sim e^{-m_{\text{scr}}z}$, must \textit{decrease} with temperature, in contrary to physical expectations. To visualise how this screening mass behaviour changes as a function of $(am_{0},g_{0})$, in Fig.~\ref{amg0Screen} we plot for $N_{s}=16$ the function
\begin{align}
\Delta m_{\text{scr}} = \frac{\frac{g_{0}}{2}J_{1}(am_{0};N_{s},N_{\tau}=1) -\frac{g_{0}^{2}}{4}J_{1}(am_{0}; N_{s},N_{\tau}=1)J_{2}(am_{0}; N_{s},N_{\tau}=1)}{\frac{g_{0}}{2}J_{1}(am_{0};N_{s},N_{\tau}=\infty) -\frac{g_{0}^{2}}{4}J_{1}(am_{0}; N_{s},N_{\tau}=\infty)J_{2}(am_{0}; N_{s},N_{\tau}=\infty)}-1,
\label{deltaM}
\end{align}
together with the bare parameter values used in the $N_{s}=16$ lattice simulations. \\

\begin{figure}[t!]
\centering
\includegraphics[width=0.5\textwidth]{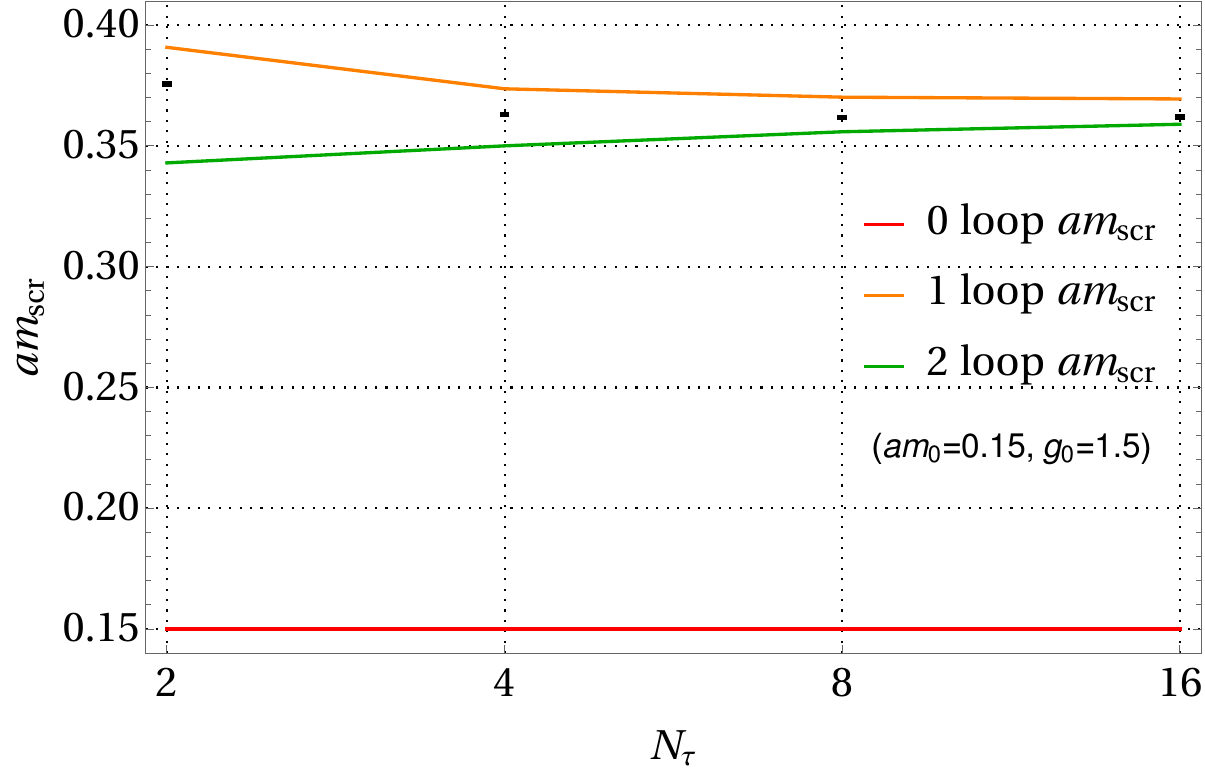}
\caption{Plot of the perturbative predictions for $am_{\text{scr}}$ compared with the fitted values from the lattice data (black points) at different values of $N_{\tau}$ for the $(am_{0}=0.15,g_{0}=1.5)$ parameter set. These values are displayed in Table~\ref{L16_Mscr}. The predictions are joined together for ease of comparison.}
\label{Mscr_am015}
\end{figure}

\noindent
When $\Delta m_{\text{scr}}<0$, and the sunset diagram contribution is negligible, the vacuum mass is larger than the screening mass at $N_{\tau}=1$, the highest possible lattice temperature of the system. From Table~\ref{tab:TL16} one can see that for the bare parameters which are closer in proximity to the region $\Delta m_{\text{scr}}<0$, as indicated by the black points in Fig.~\ref{amg0Screen}, the perturbative predictions are successively worse since $m_{\text{scr}}$ approaches a behaviour which is qualitatively opposite to that described by the data. This can be seen explicitly in Fig.~\ref{Mscr_am015} for the $(am_{0}=0.15,g_{0}=1.5)$ parameter set, where the two-loop $m_{\text{scr}}$ prediction decreases monotonically as the temperature increases. \\

\noindent
From the definitions of $J_{n}(am_{0};N_{s},N_{\tau})$ in Eq.~\eqref{Jn} it is clear that the proximity in the $(am_{0},g_{0})$-plane to the region $\Delta m_{\text{scr}}<0$ depends on the structure of the propagators entering the loop calculations. In Fig.~\ref{amg0Screen} one can see that for fixed values of $g_{0}$ one will move into this region if $am_{0}$ is sufficiently small. This can be understood as follows: since the propagators in the sums are those of a free scalar field, when $am_{0}$ becomes increasingly smaller the zero-mode ($p=0$) contribution to $J_{1}$ and $J_{2}$ will begin to dominate, leading to the behaviour
\begin{align}
J_{1}(am_{0};N_{s},N_{\tau}) \sim \frac{1}{N_{s}^{3}N_{\tau}}\frac{1}{(am_{0})^{2}}, \ \ J_{2}(am_{0};N_{s},N_{\tau}) \sim \frac{1}{N_{s}^{3}N_{\tau}}\frac{1}{(am_{0})^{4}}.
\end{align}
From this behaviour one can see that the cactus diagram contribution in the numerator of Eq.~\eqref{deltaM} will start to compete with the tadpole diagram below some critical value of $am_{0}$, and this ultimately pushes $\Delta m_{\text{scr}}$ towards negative values. The deviations seen in the two-loop predictions are therefore a consequence of the inappropriate choice of free-field propagators in the perturbative expansion. This conclusion is consistent with the issues raised in Sec.~\ref{intro}, namely that finite-temperature perturbation theory breaks down if one chooses the basic thermal field propagators to coincide with those of the vacuum theory.

\subsection{High temperature, resummations, and gauge theories \label{sec:resum}}

We have seen that perturbation theory loses accuracy as soon as temperature effects are noticeable, and that its convergence properties are qualitatively altered once $T/m\sim\mathcal{O}(1)$. Within $\phi^{4}$ theory the observed problem can only worsen as $T/m$ becomes larger, since in this case the running coupling increases with the energy scale. In the high-temperature regime it is well known that perturbation theory suffers from poor convergence properties, which has motivated the introduction of various resummation schemes~\cite{Andersen:2004fp}. These schemes are based on the reorganisation of the perturbative series via the infinite sum of certain classes of diagrams, as first explored in Ref.~\cite{Dolan:1973qd}. It has been shown that this reorganisation can be implemented iteratively by replacing the propagators in the loop calculations with the one-loop corrected ones. In the context of our results, such high-temperature resummation schemes are not applicable since we are restricted to regimes of relatively small temperatures $0<T/m\sim \mathcal{O}(1)$, which are infrared safe and do not introduce parametric mixing of couplings for soft momenta. Moreover, on a fundamental level such resummation schemes still suffer from the constraints imposed by the NRT theorem, which has been shown explicitly to result in inconsistencies in the perturbative procedure~\cite{Weldon:2001vt}. This is perhaps not surprising, since reorganisations of the perturbative series do not intrinsically modify the analytic structure of the free-field propagators used in the diagrams being resummed. \\ 

\noindent
The same question regarding propagators with real dispersion relations also occurs in all gauge theories entering the Standard Model. One might hope that for non-abelian theories, because of asymptotic freedom, the running gauge coupling at high temperatures/densities is sufficiently small to be insensitive to the issues discussed here, as was the case for the weakest coupling in Sec.~\ref{sec:fT}. However, that would be incompatible with the expected qualitative change of dynamics due to temperature/density in those cases. In any case, a high-precision lattice study of the $\mathrm{SU}(3)$-Yang-Mills equation of state at large temperatures~\cite{Giusti:2016iqr} observed that the Stefan-Boltzmann limit is approached with a slope vastly different from the leading-order perturbative predictions, and a (fitted) $\mathcal{O}(g^{6})$ contribution at $T\sim 230 \, T_{c}$ amounts to roughly 50\% of all other contributions, which suggests that similar problems in resummed perturbative schemes also occur in that theory.

\section{Addressing the issues of perturbation theory}
\label{PT_issues}

\subsection{The non-perturbative spectral function}

In Refs.~\cite{Bros:1992ey,Buchholz:1993kp,Bros:1995he,Bros:1996mw,Bros:2001zs,Bros:2003zs} the authors set out a non-perturbative framework for describing scalar QFTs at finite temperature. An important consequence of this framework is that thermal correlation functions satisfy spectral representations, which generalise the well-known K\"{a}ll\'{e}n-Lehmann representations that exist in vacuum QFTs~\cite{Kallen:1952zz,Lehmann:1954xi}. In particular, the spectral function $\rho(\omega,\vec{p})$, which is the Fourier transform of the thermal commutator $\langle [\phi(x),\phi(0)]\rangle_{\beta}$, has the representation~\cite{Bros:1996mw} 
\begin{align}
\rho(\omega,\vec{p}) = \int_{0}^{\infty} \! ds \int \! \frac{d^{3}\vec{u}}{(2\pi)^{2}} \ \epsilon(\omega) \, \delta\!\left(\omega^{2} - (\vec{p}-\vec{u})^{2} - s \right)\widetilde{D}_{\beta}(\vec{u},s).    
\label{commutator_rep}
\end{align}
Equation~\eqref{commutator_rep} implies that the dynamical effects of the thermal medium are entirely encoded in the \textit{thermal spectral density} $\widetilde{D}_{\beta}(\vec{u},s)$. Besides the general analytic constraints satisfied by the thermal spectral density, in Ref.~\cite{Bros:1992ey} it was proposed that the thermal medium could contain particle-like excitations, and that in position space these would contribute to $D_{\beta}(\vec{x},s)$ as discrete components of the form: $D_{m,\beta}(\vec{x})\, \delta(s-m^{2})$, where $m$ is the mass of a vacuum particle state. Since $D_{m,\beta}(\vec{x})\rightarrow 1$ in the zero-temperature limit, due to the restoration of Lorentz symmetry, these components represent a finite-temperature generalisation of stable vacuum states. In order to draw a sharp distinction with other thermal excitations considered in the literature, such as collective quasi-particle modes which vanish when $T\rightarrow 0$, these contributions were subsequently referred to as \textit{thermoparticles}~\cite{Buchholz:1993kp}. \\

\noindent
Given a system with a single $T=0$ particle species of mass $m$, it was proposed in Ref.~\cite{Bros:2001zs} that the thermal spectral density has the decomposition     
\begin{align}
D_{\beta}(\vec{x},s)= D_{m,\beta}(\vec{x})\, \delta(s-m^{2}) + D_{c, \beta}(\vec{x},s),
\label{decomp}
\end{align}
where $D_{c, \beta}(\vec{x},s)$ parametrises all other spectral contributions. As summarised recently~\cite{Lowdon:2022xcl,Lowdon:2022yct,Bala:2023iqu}, there are several well-motivated reasons for why thermoparticles are a natural description for what happens to particle states in the presence of a thermal medium. Perhaps the most compelling is that on account of the representation in Eq.~\eqref{commutator_rep}, the appearance of a non-trivial coefficient $D_{m,\beta}(\vec{x})$ necessarily causes the zero-temperature peak in $\rho(\omega,\vec{p})$ at $p^{2}=m^{2}$ to become broadened, which describes the effect of collisions with the medium. In position space this is reflected in the appearance of the multiplicative term $D_{m,\beta}(\vec{x})$ which inhibits the propagation of these states, i.e. lowers their mean-free path, and hence represents a thermal damping factor.

\subsection{Thermoparticles in $\phi^{4}$ theory}

We now investigate if the $\phi^{4}$ theory lattice data analysed in this study is consistent with the presence of thermoparticle-like excitations. In particular, we focus on the $(am_{0}=0.15,g_{0}=1.5)$ parameter set for $N_{s}=16$, since this is where the largest temperature-dependent deviations were seen with the standard perturbative predictions. For this purpose we follow the approach set out in Refs.~\cite{Bala:2023iqu,Lowdon:2022xcl}, where QCD correlator data of light pseudo-scalar mesons was used to detect the presence of thermoparticle excitations at low energies. In particular, we perform the following steps:

\begin{enumerate}

\item Perform a fit of the spatial correlator $C(z)$ at each temperature. The data is consistent across all $z$ data points with an exponential behaviour, and so we fit the following functional form:
\begin{align}
C(z) = A \, e^{-m_{\text{scr}}z} + A \, e^{-m_{\text{scr}}(aN_{s}-z)}, 
\label{Cz_fit}
\end{align}
where $m_{\text{scr}}$ is the screening mass and the second term is included in order to take into account the spatial periodic boundary conditions of the fields. The non-perturbative vacuum mass $m$ is estimated from the smallest temperature sample of the parameter set, namely: $m= m_{\text{scr}}(N_{\tau}=16)$. The results of these fits are listed in Table~\ref{L16_Mscr}.

\begin{table}[ht!]
\center
\small
\renewcommand{\arraystretch}{1.16}
\begin{tabular}{|c|c|c|c|c|} 
\hline
\rule{0pt}{3ex}
$N_{s}^{3} \times N_{\tau}$ &  $(am_{0}, g_{0})$ &  $A$ &  $am_{\text{scr}}$    & $\chi^{2}/\text{d.o.f.}$  \\[0.5ex]
\hhline{|=|=|=|=|=|}
$16^{3} \times 16$     & $(0.15, 1.5)$   &  1.355(2) &  0.3620(6)   &  0.40     \\
\hline
$16^{3} \times 8$      & $(0.15, 1.5)$   &  1.350(1) &  0.3618(2)   &  0.05     \\
\hline
$16^{3} \times 4$      & $(0.15, 1.5)$   &  1.347(1) &  0.3629(3)	&  0.06     \\
\hline 
$16^{3} \times 2$      & $(0.15, 1.5)$   &  1.301(2) &  0.3757(7)   &  0.55     \\ 
\hline
\end{tabular}
\caption{The fitted parameter values using the fit ansatz in Eq.~\eqref{Cz_fit} for $N_{s}=16$ with lattice parameters $(am_{0}=0.15,g_{0}=1.5)$.}
\label{L16_Mscr}
\end{table} 

\item From the analysis in Refs.~\cite{Bala:2023iqu,Lowdon:2022xcl} the appearance of the components in Eq.~\eqref{Cz_fit} are consistent with the presence of a single thermoparticle state with damping factor
 \begin{align}
D_{m,\beta}(\vec{x}) =  \alpha \, e^{-\gamma|\vec{x}|}, \quad\quad \alpha = 2 A \, am_{\text{scr}}, \  \gamma= m_{\text{scr}}-m,
\label{thermo_damping_scalar}
\end{align} 
where $\gamma \geq 0$. The contribution of this thermoparticle state to the spectral function $\rho(\omega,\vec{p})$ has the analytic form
\begin{align}
\rho_{\text{TP}}(\omega,\vec{p}) = \epsilon(\omega)  \theta(\omega^{2}-m^{2}) \,  \frac{4 \, \alpha \gamma  \sqrt{\omega^{2}-m^{2}}}{(|\vec{p}|^{2}+m^{2}-\omega^{2})^{2} + 2(|\vec{p}|^{2}-m^{2}+\omega^{2})\gamma^{2}+\gamma^{4} }. 
\label{thermo_spec} 
\end{align}
A characteristic feature of Eq.~\eqref{thermo_spec} is that both the interaction and temperature dependence are entirely contained in the thermal width-like parameter $\gamma$.

\item Restricting to $\vec{p}=0$, Eq.~\eqref{thermo_spec} and the fitted parameters $(m_{\text{scr}}, \gamma, m)$ can then be used to compute the prediction of the \textit{temporal} correlator at zero momentum\footnote{On the lattice this correlator is defined: $\widetilde{C}(\tau,\vec{p}=0) =  a^{3}\sum_{\tau,x,y}\langle \phi((\tau,\vec{x})\phi(0)\rangle$.}, and compared with the corresponding lattice data. This provides a non-trivial check of whether Eq.~\eqref{thermo_spec} gives a consistent description of the spectral function. 

\end{enumerate}

\begin{figure}[t!]
\centering
\includegraphics[width=0.49\textwidth]{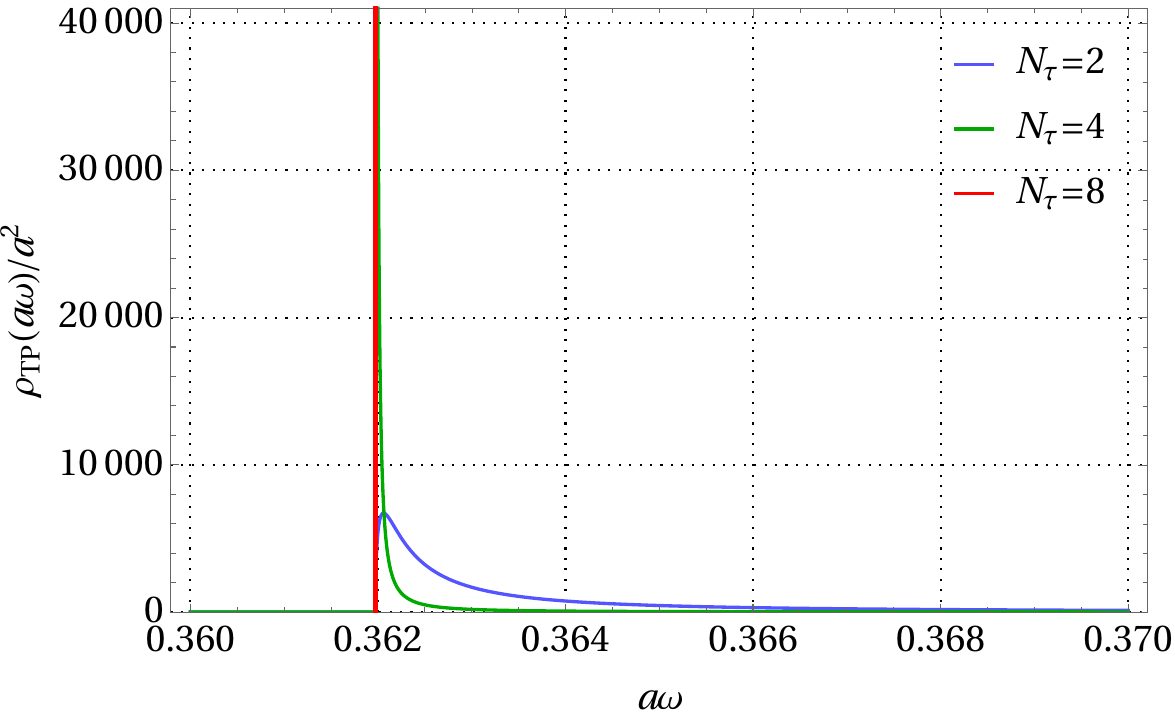}
\includegraphics[width=0.46\textwidth]{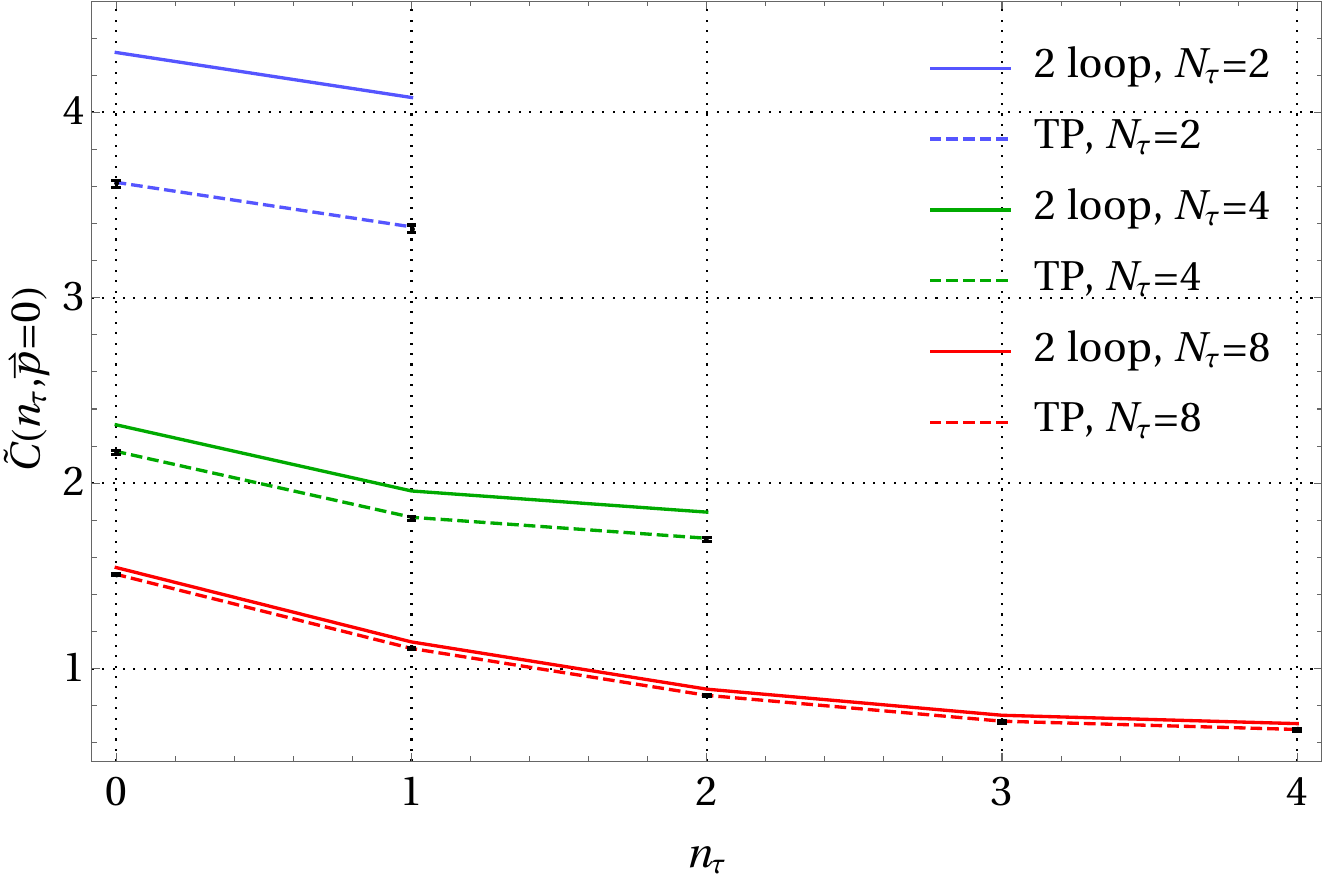}
\caption{Lattice temporal correlator data and predictions for $N_{s}=16$ and $(am_{0}=0.15,g_{0}=1.5)$ (right). The solid lines indicate the two-loop predictions from lattice perturbation theory, and the dashed lines are the predictions using a thermoparticle (TP) spectral function, the parameters of which are extracted from the spatial correlator data. The form of the corresponding TP spectral functions are displayed in the left plot for each temperature.}
\label{CtPred16}
\end{figure}

\noindent
The results from applying the procedure set out in steps 1-3 are shown in Fig.~\ref{CtPred16}. In the left panel is displayed the form of the zero-momentum thermoparticle spectral function for each non-trivial temperature, and in the right we compare the predictions from standard two-loop lattice perturbation theory (solid lines) and those using these spectral functions (dashed lines). The thermoparticle prediction is consistent with the lattice data within statistical errors, which is clearly an improvement over the perturbative results, since these deviate further away for larger temperatures. In general, the full spectral function of the theory should satisfy the sum rule
\begin{align}
\int_{-\infty}^{\infty} \frac{d\omega}{2\pi} \omega \, \rho(\omega,\vec{p}) = 1. 
\end{align}
It therefore follows from the form of the thermoparticle component $\rho_{\text{TP}}(\omega,\vec{p})$ in Eq.~\eqref{thermo_spec} that    
\begin{align}
\int_{-\infty}^{\infty} \frac{d\omega}{2\pi} \omega \, \rho_{\text{TP}}(\omega,\vec{p}) = \alpha < 1,
\label{spec_bound}
\end{align}
and hence $\alpha$ describes the relative spectral weight of the thermoparticle component. Using the relation for $\alpha$ in Eq.~\eqref{thermo_damping_scalar}, together with the extracted parameter values in Table~\ref{L16_Mscr}, one finds that $\alpha$ does indeed satisfy the bound in Eq.~\eqref{spec_bound} at each temperature, and that this bound is almost saturated. Overall, these findings suggest that similarly to the QCD spectral functions studied in Refs.~\cite{Bala:2023iqu,Lowdon:2022xcl}, the lattice $\phi^{4}$ theory data is consistent with the presence of a thermoparticle component. However, in contrast to the QCD case, where higher excited states also provide a significant contribution, here the lowest thermoparticle component entirely dominates the spectral function at all temperatures considered. This confirms that thermoparticle excitations provide an important non-perturbative spectral contribution in $\phi^{4}$ theory, particularly at low temperatures.

\subsection{Thermoparticle perturbation theory}
\label{TP_PT}

\noindent
There are several good reasons to believe that thermoparticles could also play a central role in finite-temperature perturbation theory:  

\begin{itemize}

\item Given the decomposition in Eq.~\eqref{decomp} it has been proven~\cite{Bros:2001zs} that these components \textit{dominate} the large real-time $x_{0}$ behaviour of thermal correlation functions, just as stable particle states do at zero temperature. Thermoparticles therefore satisfy a required property of any would-be finite-temperature scattering state. By analogy to the vacuum case, these states should ultimately form the basis of any perturbative expansion, and hence the basic field propagators appearing in the perturbative diagrams would be those of the thermoparticle states.

\item The appearance of the damping factor $D_{m,\beta}(\vec{x})$ prevents thermoparticles from having a delta function-like spectral function with a sharp dispersion law, and hence these states avoid the constraints imposed by the NRT theorem. This therefore opens up the possibility of having a non-trivial finite-temperature $S$-matrix.

\item In Ref.~\cite{Bros:2001zs} a consistency condition was proposed for thermal fields at large times which encodes the asymptotic dynamics of the theory. It was demonstrated that if thermoparticles describe thermal asymptotic states, then the form of their corresponding damping factors, and hence propagators, are \textit{uniquely} fixed by this condition. Ultimately, this suggests that the finite-temperature propagators that should be used in any perturbative QFT expansion can be self-consistently derived from that QFT.  

\end{itemize}

\noindent
Taken together, these characteristics suggest that the propagators associated with thermoparticle states potentially provide the correct basis for performing perturbative expansions at finite temperature, and that the resulting \textit{thermoparticle perturbation theory} framework could be used to obtain consistent finite-temperature perturbative predictions. This question has been explored before in Ref.~\cite{Bros:2001zs}, but it remains to be seen whether such an expansion possesses all of the correct characteristics, including renormalisability. It would also be interesting to understand how the dominance of the thermoparticle component changes as the temperature of the system is raised further. Such an analysis would in principle be able to establish the relative interplay between the different spectral components in Eq.~\eqref{decomp}, and is left for future work.

\section{Conclusions}
\label{concl}

Finite-temperature QFTs are subject to constraints that are no longer present in the vacuum formulation of these theories. One of the most consequential such constraints is the inability to construct interacting thermal states with purely real dispersion relations. This is the implication of the Narnhofer-Requardt-Thirring (NRT) theorem. From a perturbative context, this implies that neither free field, nor resummed quasi-particle-like propagators with real poles, can form the basis of a finite-temperature perturbative expansion. In this work we set out to test this constraint by comparing the standard perturbative predictions of scalar correlation functions in $\phi^{4}$ theory with the results from lattice simulations. For this purpose we used lattice perturbation theory so that a direct comparison could be made with the lattice data without having to address the known subtleties associated with the continuum limit of the theory. By computing spatial correlator predictions up to two-loop order using a range of different volumes, temperatures, and bare lattice parameters, we find that these predictions deteriorate with increasing temperature, and that the deviations are a direct result of the analytic structure of the free-field propagators used in the perturbative expansion. These deviations reflect the fundamental constraints imposed by the NRT theorem, which go beyond the infrared regime of the theory. This is consistent with the conclusions of Ref.~\cite{Weldon:2001vt}, where the finite-temperature perturbative approach is shown to explicitly break down in $\phi^{4}$ theory at two-loop order. \\

\noindent
In the remainder of this work, we used the non-perturbative QFT framework set out in Refs.~\cite{Bros:1992ey,Buchholz:1993kp,Bros:1995he,Bros:1996mw,Bros:2001zs,Bros:2003zs} to investigate how these issues can potentially be resolved. An important consequence of this framework is the identification of distinct particle-like thermal excitations, so-called thermoparticles. By analysing both spatial and temporal correlator data we find evidence for the existence of a single such thermoparticle component, and that this dominates the spectral function at the low temperatures considered. These findings suggest that thermoparticles are the basic constituents of the thermal medium in $\phi^{4}$ theory, and hence any consistent perturbative expansion should ultimately be parametrised in terms of these non-perturbative degrees of freedom.

\section*{Acknowledgements}

The authors would like to thank J.-P.~Blaizot, D.~B\"{o}deker, R.~Pisarski, and D.~Rischke for useful discussions and input. The authors acknowledge support by the Deutsche Forschungsgemeinschaft (DFG, German Research Foundation) through the Collaborative Research Center CRC-TR 211 ``Strong-interaction matter under extreme conditions'' -- Project No. 315477589-TRR 211. O.~P.~also acknowledges support by the State of Hesse within the Research Cluster ELEMENTS (Project ID 500/10.006).

\appendix

\section{Lattice simulations}
\label{lattice_details}

For the purpose of the numerical simulations in this work, the lattice action in Eq.~\eqref{eq:latact1} is rewritten in terms of dimensionless fields as follows
\begin{align}
&S = \sum_{x \in \Lambda_{a}} \left[ - 2\kappa \sum_{\mu} \varphi(x) \varphi(x+ a \hat{n}_{\mu}) + \varphi(x)^{2} + \lambda [ \varphi(x)^{2} -1 ]^{2} \right], \\
&a\phi(x) = (2 \kappa)^{\frac{1}{2}} \varphi(x), \quad  a^{2}m_{0}^{2} = \frac{1 - 2 \lambda}{\kappa} - 8, \quad g_{0} = \frac{6 \lambda}{\kappa^{2}}.
\end{align}
Our Monte Carlo simulations were based on a heat bath algorithm for the Gaussian part of the action, supplemented by a Metropolis step for the quartic term. In order to maximise decorrelation we have used a mixture of heatbath and reflection updates, as suggested in Ref.~\cite{Bunk:1994xs} for an $\mathrm{SU}(2)$-Higgs model, which we adapted to trivial gauge fields and a one-component scalar field. One compound sweep then consisted of three reflection steps per heat bath update. For the $N_{s}=4$, 16, and 64 lattices we collected 400k, 100k, and 60k-100k field configurations, respectively. All errors were computed by a jackknife analysis and explicitly checked to be stable under a change of the bin size by factors of up to 10, with $>60$ bins always available in the end.

\bibliographystyle{JHEP}

\bibliography{refs}

\end{document}